\documentclass{book}    
\usepackage{piers}  
\begin{document}

\title{Near-field properties of plasmonic nanostructures with high aspect ratio}
\maketitle


\begin{authors}
{\bf Y. Ould Agha}$^{1}$, {\bf O. Demichel}$^{1}$, {\bf C. Girard}$^{2}$ {\bf A. Bouhelier}$^{1}$ and {\bf G. {Colas des Francs}}$^{1}$\\
\medskip
$^{1}$Laboratoire Interdisciplinaire Carnot de Bourgogne (ICB) \\
UMR 6303 CNRS/Universit\'e de Bourgogne\\
9, Av. Savary, BP 47870, 21078 Dijon Cedex, France\\
$^{2}$Centre d'Elaboration de Mat\'eriaux et d'Etudes Structurales (CEMES) \\
CNRS, 29 rue J. Marvig F-31055 Toulouse, France
\end{authors}


\begin{paper}

\begin{piersabstract}
Using the Green's dyad technique based on cuboidal meshing, 
we compute the electromagnetic field scattered by metal nanorods with high aspect ratio. We investigate the effect of the meshing shape on the numerical simulations. We observe that discretizing the object with cells with aspect ratios similar to the object's aspect ratio improves the computations, without degrading the convergency. 
We also compare our numerical simulations to finite element method and discuss further possible improvements. 
\end{piersabstract}

\psection{Introduction}
Plasmonic nanostructures concentrate light on subwavelength scales \cite{Agio:2012,Derom-Vincent-Bouhelier-GCF:2012}, 
opening the way to promising applications such as nano-optical antennas 
\cite{Bharadwaj-Deutsch-Novotny:2009,Olmon-Raschke:2012}, 
or surface enhanced spectroscopies \cite{Moskovits:1985,Pettinger:2010}. 
In the last decade, important efforts have been done in modeling these 
structures \cite{Girard2005}.  
Particularly, the electric field rapidly decays away from the metal surface so that dedicated methods have to be developed to correctly 
describe near-field optics properties of plasmonic systems. 
Additionnally, high aspect ratio shapes are often needed to tune resonance frequencies \cite{Marzan:2008,Dujardin:2013}. 
Fast field decay and high aspect ratio lead to serious numerical difficulties for methods based on the discretization of the object volume, 
notably in term of memory cost and computational time. 

We are specifically interested in the Green's dyad technique (GDT) (also called volume integral method) \cite{lakhtakia92} 
since it naturally satisfies boundary conditions 
both at the object surfaces and in the far-field. Moreover, it easily includes a substrate supporting nanoparticles 
without the need to discretize it \cite{Paulus2001}. In addition, as far as the Green's dyadic is numerically computed in the source 
(nanostructures) region, all the electromagnetic properties of the system are easily obtained with practically 
almost no additionnal computing cost \cite{PRELevequeGCF:2002}. 
This concerns obviously the electric and magnetic fields in the very near-field of the object \cite{Girard-Weeber-Dereux-Martin-Goudonnet:1997} 
but also scattering properties in the radiation zone \cite{Novotny1997b,CPLGCF:2001}. Last, the local density of optical states (LDOS), 
an intrinsic property of the  nanostructure,  is also easily deduced from this formalism \cite{CPLGirardGCF:2005}. 
 
Recent works were devoted to improve the GDT when applied to metallic nanostructures. 
The GDT relies on the discretization of the scatterers. Because of strongly varying fields, 
the key point is to correctly describe the fast variation of the Green's tensor over an elementary cell.  
Let us mention the regularization scheme proposed by Kottmann and Martin developed for 2D-elements 
but also transposable to 3D-nanostructures \cite{Kottmann-Martin:2000}. Instead of regularizing the Green's tensor, Chaumet and coworkers directly 
computed its integral over the volume of cubic meshes \cite{Chaumet-Sentenac-Rahmani:2004}. To this aim, they considered a Weyl expansion 
of the tensor and performed a numerical integration in the reciprocical k-space. 
In the present work, we first transform the volume integral to a surface integral over a cuboidal mesh \cite{Gao-TorresVerdin-Habashy:2005} 
before a numerical evaluation in direct space. This avoids the singularity at the mesh center and convergence difficulties. 
Moreover, this permits to consider more complex meshing. Note finally that our method differs from surface integral method (also called boundary element method) where 
the \emph{object} surface is discretized, including the substrate if necessary \cite{Myroshnychenko-GarciaAbajo:2008,Hohenester-Trugler:2008,Kern-Martin:2009}. Surface integral method relies on the introduction of equivalent surface currents that depends on the excitation field. Differently, the volume integral method relies on a volume discretization of the object, excluding the substrate, but in the present work involves integration over the surface of the meshes. This leads to the evaluation of the Green's dyad associated to the whole structure that is independent on the illumination conditions so that various problem can be treated by post-processing. 

In the following, we define a benchmark configuration that consists of a silver nanorod. 
In order to assess the reliability of our results, we compare our numerical simulations 
to those obtained using a commercial software (COMSOL Multiphysics) based on the finite element method (FEM) \cite{comsol}. 
The objective of this comparison is twofold. FEM efficiently considers complex shape but necessitates a careful design and positioning 
of a perfectly matched layer (PML) to avoid artefact reflexion at the boundaries of the computational window. As we will see, it becomes a very sensitive parameter  in presence of a substrate.  On the opposite, GDT easily considers nanostructures deposited on a substrate but is generally limited to simpler shapes.  
By comparing the two methods, we are able to estimate the error. Reciprocally, this validates the choice of the PML included in the FEM calculations. 

\psection{Numerical methods}
\label{sect:theory}

\psubsection{Green's dyadic technique}

We consider an object immersed in a homogeneous background medium. 
The addition of a substrate will be discussed later. 
The object and background are assumed nonmagnetic ($\mu=1$) with a relative permittivity, $\epsilon$ and $\epsilon_B$, respectively.  
In the following, we assume an $\exp(-i\omega t)$ time harmonic dependence for the fields. 
The Green's dyad represents the electromagnetic response to an elementary excitation. Physically, the electric field 
scattered at the position ${\bf r}$ in presence of a point-like dipolar source ${\bf p_0}$ located at ${\bf r_0}$ follows

\begin{equation}
{\bf E}({\bf r})=\frac{k_0^2}{\epsilon_0}{\bf G}({\bf r},{\bf r_0}) \cdot {\bf p_0} \,,
\label{eq:LippSchwin}
\end{equation}
where $\epsilon_0$ is the free-space permittivity, $k_0$ the free-space wavenumber and 
${\bf G}({\bf r},{\bf r_0})$ is the Green's tensor associated with the whole system. 
It is an intrinsic electromagnetic quantity of the object, 
independent on the excitation process. For instance the local density of mode is given by \cite{PRLChicanneGCF:2001,JCPGCF:2002}
\begin{equation}
\rho({\bf r},\omega)=\frac{\omega}{\pi c^2}Im[{\bf G}({\bf r},{\bf r})] \,.
\label{eq:LDOS}
\end{equation}
In addition, if the object is excited by an incident electric field ${\bf E_0}$, the electric field  can be expressed everywhere in the system by a 3D integral
\begin{equation}
{\bf E}({\bf r})={\bf E_0}({\bf r})+k_0^2\iiint_{object}{\bf G}({\bf r},{\bf r'}) 
\cdot  \Delta \epsilon {\bf E_0}({\bf r'}) d{\bf r'}\,, 
\label{eq:LippSchwin}
\end{equation}
with $\Delta \epsilon=\epsilon({\bf r})-\epsilon_B$.
Finally, the main numerical task is the evaluation of the tensor ${\bf G}$ for every couple of points $({\bf r},{\bf r'})$. 
The Green's tensor can be computed by solving the self-consistent Dyson's equation
\begin{eqnarray}
{\bf G}({\bf r},{\bf r'})={\bf G_0}({\bf r},{\bf r'})+k_0^2\iiint_{object}{\bf G_0}({\bf r},{\bf r''}) \cdot
\Delta \epsilon \cdot {\bf G}({\bf r''},{\bf r'}) d{\bf r''} \,, 
\label{eq:Dyson}
\end{eqnarray}
where ${\bf G_0}$ is the free-space Green's dyad, that is analytical. In order to solve Dyson's equation (\ref{eq:Dyson}), 
the object is discretized into $N$ cells of volume $V_k$ ($k=1,\cdots, N$); 
\begin{eqnarray}
\label{eq:DysonDisc}
&&{\bf G}({\bf r},{\bf r'})={\bf G_0}({\bf r},{\bf r'})+k_0^2\sum_{k=1}^N{\bf G_0^{int}}({\bf r},{\bf r_k})\cdot
\Delta \epsilon \cdot {\bf G}({\bf r_k},{\bf r'})  \,, \text{with} \\
&&{\bf G_0^{int}}({\bf r},{\bf r_k})=\iiint_{V_k}{\bf G_0}({\bf r},{\bf r''})d{\bf r''} \,. 
\label{eq:Gint}
\end{eqnarray}
Depending on the object shape, different meshes can be used. In case of spherical cells, 
the integral of the Green's tensor over a sphere of radius $a_k$, centered at ${\bf r}_k$ 
is analytical and writes \cite{Gao-TorresVerdin-Habashy:2005}

\begin{eqnarray}
\label{eq:GintSphere}
&&{\bf G_0^{int}}({\bf r},{\bf r_k})=C_k{\bf G_0}({\bf r},{\bf r_k}) ,\text{with} \\
&&C_k=\frac{4\pi a_k}{k_B^2}\left(\frac{\sin(k_Ba_k)}{k_Ba_k}-\cos(k_Ba_k) \right), \text{where} \, k_B=\sqrt{\epsilon_B}k_0\,.
\end{eqnarray}
$C_k$ is a geometrical factor that reduces to the sphere volume $C_k=4\pi a_k^3/3=V_k$ of subwavelength size ($k_Ba_k \ll 1$). 
In this case, the discretized Dyson's equation \ref{eq:DysonDisc} reduces to \cite{Girard2005,lakhtakia92}
\begin{eqnarray}
{\bf G}({\bf r},{\bf r'})={\bf G_0}({\bf r},{\bf r'})+k_0^2\sum_{k=1}^N {V_k} {\bf G_0}({\bf r},{\bf r_k}) \cdot
\Delta \epsilon \cdot {\bf G}({\bf r_k},{\bf r'})\,.  
\label{eq:DysonDisc0}
\end{eqnarray}

However, when plasmonic objects with high aspect ratio are considered, it is preferable to use elongated cells. And the volume integral Eq. \ref{eq:Gint} has to be numerically evaluated. 
This is rather difficult because of the singularity of the Green's tensor that occurs 
when the source and the observation points coincide (${\bf r'}={\bf r}$) \cite{Yaghjian:1980}. In a smilar context, 
Chaumet {\it et al} used a Weyl expansion of the Green's tensor to performed the numerical integration of  Eq. \ref{eq:Gint} in the reciprocical k-space 
\cite{Chaumet-Sentenac-Rahmani:2004}.
More recently, Massa and cowokers derived an approximate analytical expression for the polarisability of a cuboidal cell \cite{Massa-Maier:2014}.
 
In the present work, we first transform the volume integral to a surface integral over the (cuboidal) cell surface before a numerical evaluation 
in direct space. This avoids the singularity at the cell center and convergence difficulties.

In 2005, Gao {\it et al} demonstrated that this volume integral ${\bf G_0^{int}}$ 
can be converted into a (flux) surface integral thanks to Ostrogradsky's theorem \cite{Gao-TorresVerdin-Habashy:2005}

\begin{eqnarray}
\label{eq:GintS}
{\bf G_0^{int}}({\bf r},{\bf r_k})&=&\iiint_{V_k}{\bf G_0}({\bf r},{\bf r''})d{\bf r''} 
=-{\bf D}({\bf r})+\iint_{S_k}{{\bf g}_0}({\bf r},{\bf r''})d{\bf r ''}\,. 
\label{eq:GintSurf}
\end{eqnarray}
where ${\bf D}({\bf r})=1/k_B^2$ if the observation point is within the integration volume $({\bf r} \in V_k$) and is null elsewhere. 
For a rectangular parallelepiped mesh ($a \times b \times c$), centered at the point ${\bf r}_c=(x_c,y_c,z_c)$, the surface integral term writes for instance 
\begin{eqnarray}
{G_{0,xx}^{int}}({\bf r},{\bf r_c})=\frac{1}{4\pi k_B^2}\int_{y_c-b/2}^{y_c+b/2}\int_{z_c-c/2}^{z_c+c/2}
\left[
\frac{(x-x_c-a/2)e^{ik_B R_{x1}}(ik_b R_{x1}-1)}{R_{x1}^3} \right. \\
\left.
-\frac{(x-x_c+a/2)e^{ik_B R_{x2}}(ik_b R_{x2}-1)}{R_{x2}^3}
\right]dz_0 dy_0 \,. 
\label{eq:GintSurfxx}
\end{eqnarray}
with $R_{x1,2}=[x-x_c\pm a/2)^2+(y-y_0)^2+(z-z_0)^2]^{1/2}$ and $\pm$ refers to either $R_{x1}$ (plus sign) or $R_{x2}$ (minus sign). Such surface integral is efficiently numerically computed using the Gauss-Kronrod method.  
The singularity of the Green's dyad in the source region is properly taken into account in this approach. 
Another notable advantage of this approach is that it considerably reduces the memory cost for evaluating the Green's dyadic. Finally, different mesh shapes can be considered by adapting the integral surface.    
And Dyson's equation (Eq. \ref{eq:DysonDisc}) is numerically solved using standard matrix inversion techniques.

\psubsection{Finite element method}
The finite element method is a common numerical tool to study problems related to electromagnetism. Its main advantage is that it can be applied to irregular geometries. Since a detailed description can be found in Ref. \cite{Jin:2002}, only a brief introduction will be given here. 
In classical electromadynamics, FEM is typically used to solve problems governed by Maxwell's equations 
that can be formulated as a vector wave equation for the electric field
\begin{equation}
\nabla \wedge [ \frac{1}{\mu} \nabla \wedge \bf{E}]  -\mathrm{k}_{0}^{2}\epsilon \bf{E}=0\,.
\label{MaxFEM}
\end{equation}

The finite element method is based on the discretization of the whole geometry,  including the surrounding medium, by simple elements 
(such as triangles in two dimensional problems, or tetrahedral elements in three dimensional problems). 
The system of equations to be solved can be assembled after obtaining the weak formulation of the partial differential equations.


However, one of the main difficulties in the finite element analysis of wave problems in open space is to truncate the unbounded domain. 
A common approach is to introduce a special layer of finite thickness surrounding the region of interest such that it is non-reflecting 
and completely absorbing for the waves entering this layer under any incidence. Such regions were introduced by Berenger and 
are called perfectly matched layers (PML) \cite{Berenger}. The most natural way is to consider PML as a geometrical transformation that leads to equivalent 
$\epsilon$ and $\mu$ 
(that are complex, anisotropic, and inhomogeneous even if the original ones were real, isotropic, and homogeneous) \cite{Yacoub1, Yacoub2} . 
This leads automatically to an equivalent medium with the same impedance than the one of the initial ambient medium 
since $\epsilon$ and $\mu$ are transformed in the same way. The equivalent  $\epsilon$ and $\mu$ ensure that the interface with the layer is non-reflecting. 


\psection{results and discussion}
\label{sect:simu}
\begin{figure}[h!]
  \centering
 \includegraphics[width=5cm]{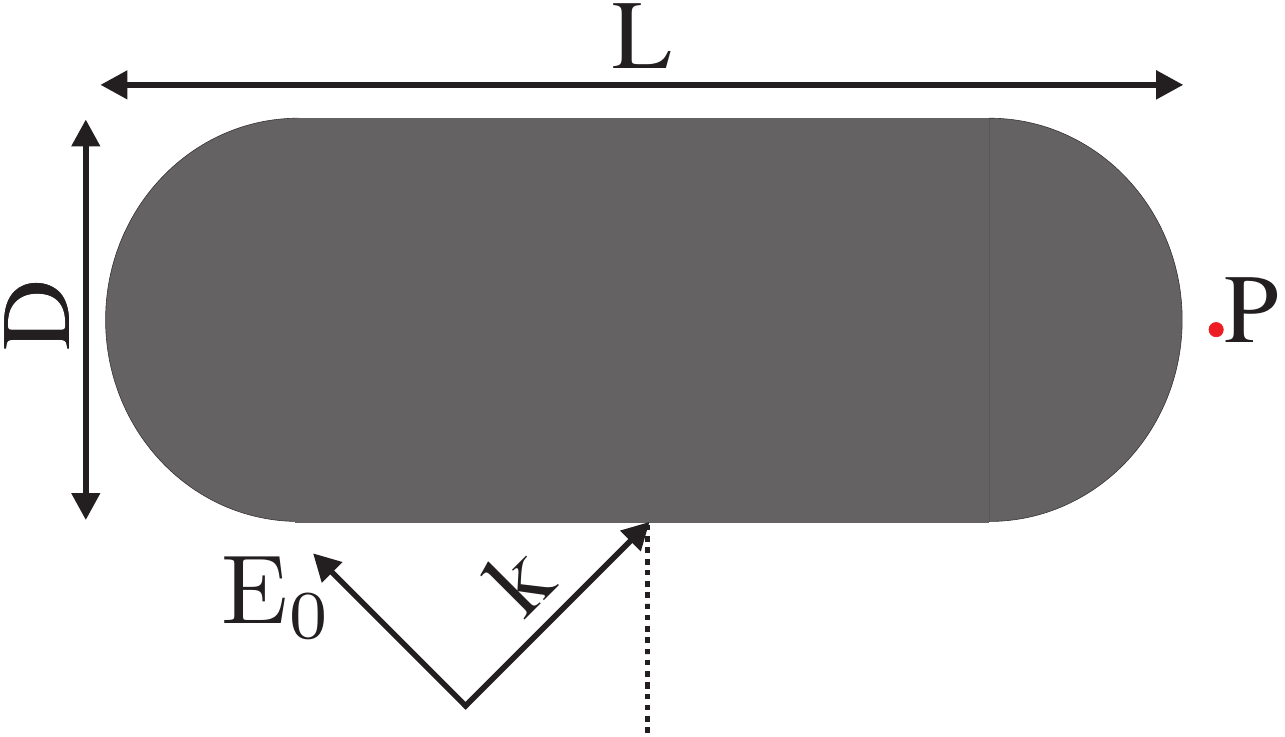}
 \caption{Benchmark problem: a silver nanorod of aspect ratio $L/D$ in air.}
\label{Monomer}
\end{figure}

In the following, we consider a silver nanorod that consists of a cylinder rod capped with hemispherical ends 
(see Fig.~\ref{Monomer}). The diameter of the nanorod is $D=2R$ and the total length is $L$ so that its aspect ratio is $L/D$.
In this part of our numerical model, the silver nanorod is immersed in air. We use the silver dielectric function tabulated by 
Johnson and Christy \cite{Johnson-Christy:1972}.

\begin{figure}[ht]
\begin{center}
  \subfigure[]{\epsfig{figure=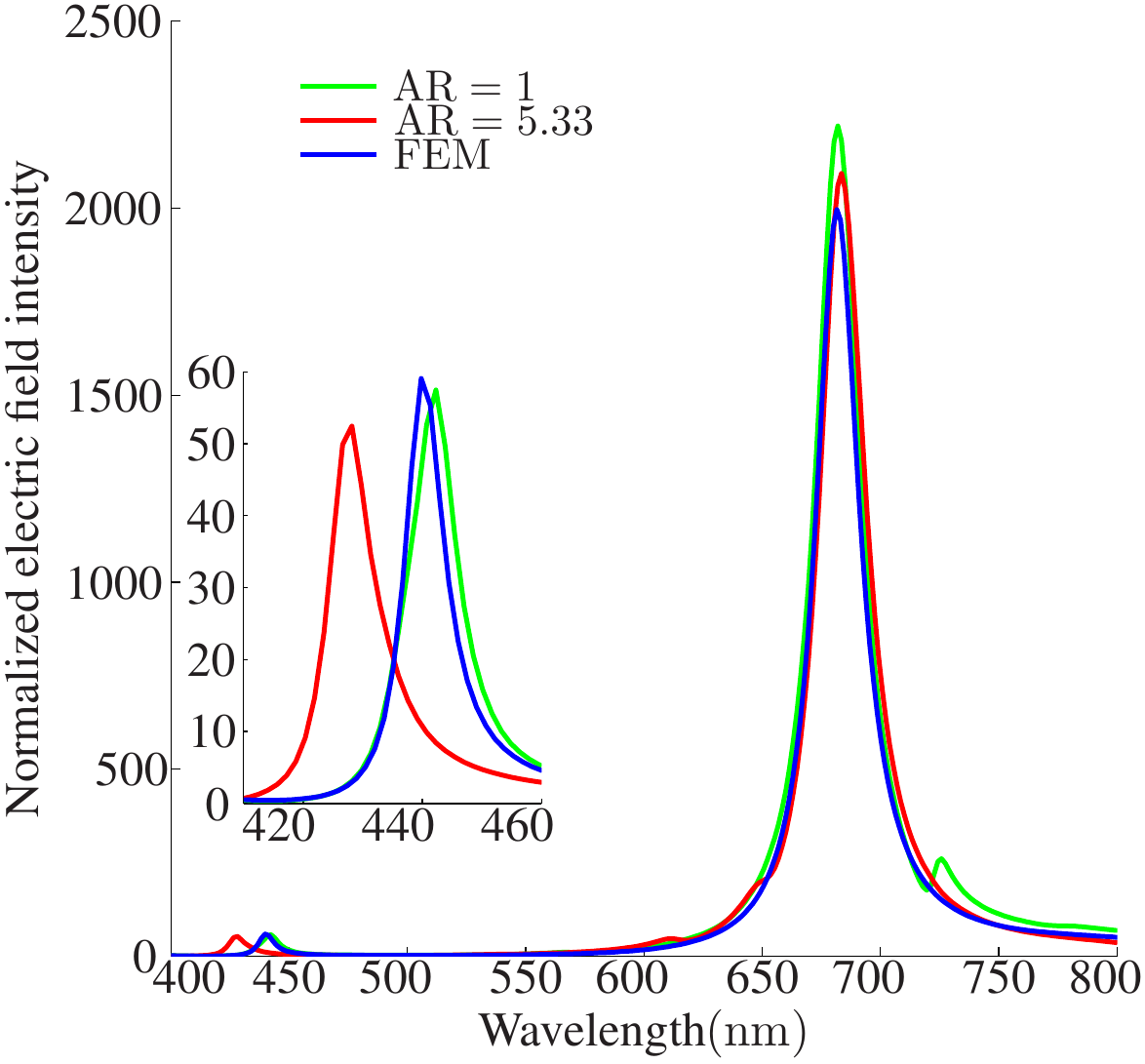,height=6cm}\label{Spectres}}\quad
  \subfigure[]{\epsfig{figure=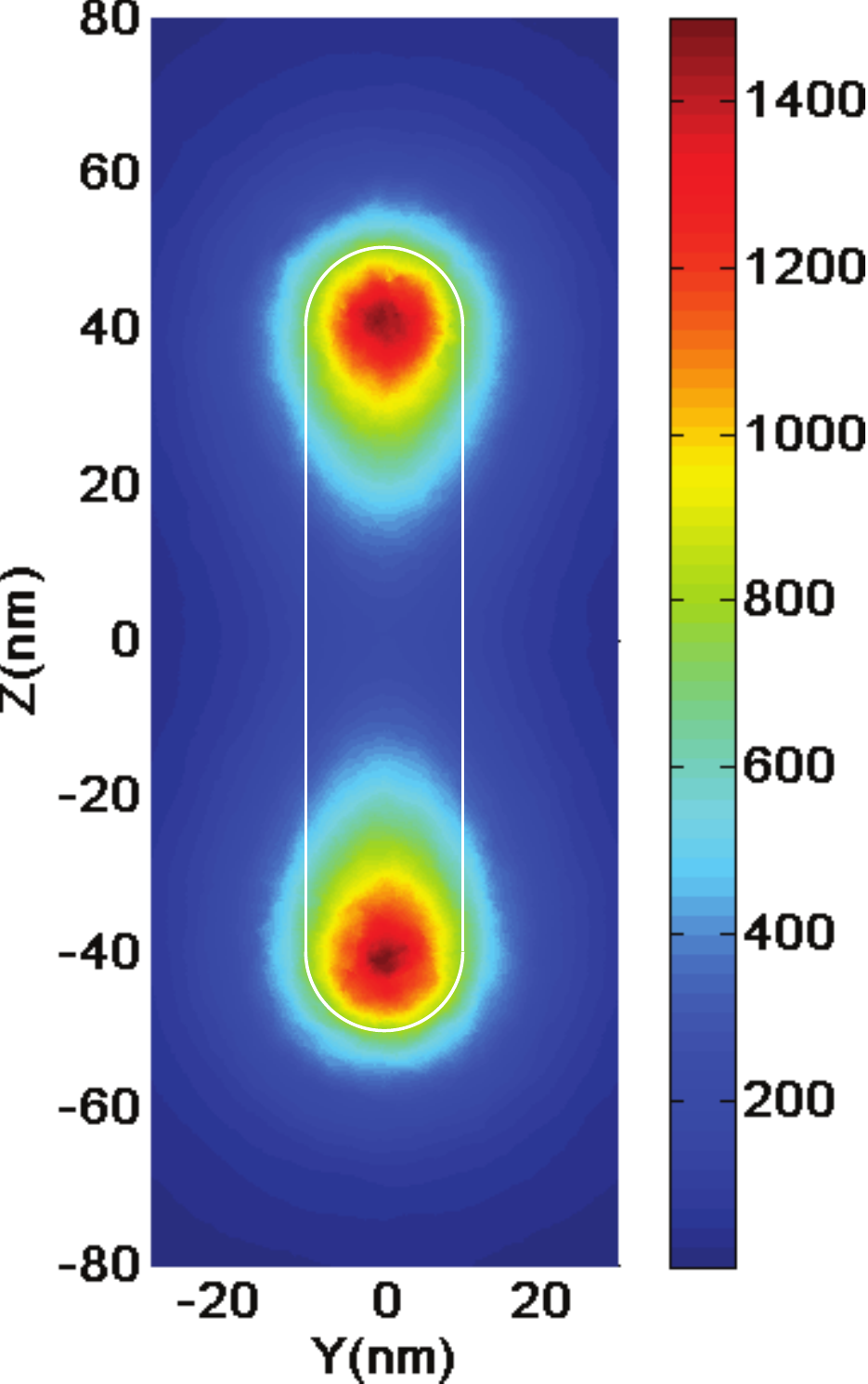,height=6cm}\label{FM_Comsol}}\quad
  \subfigure[]{\epsfig{figure=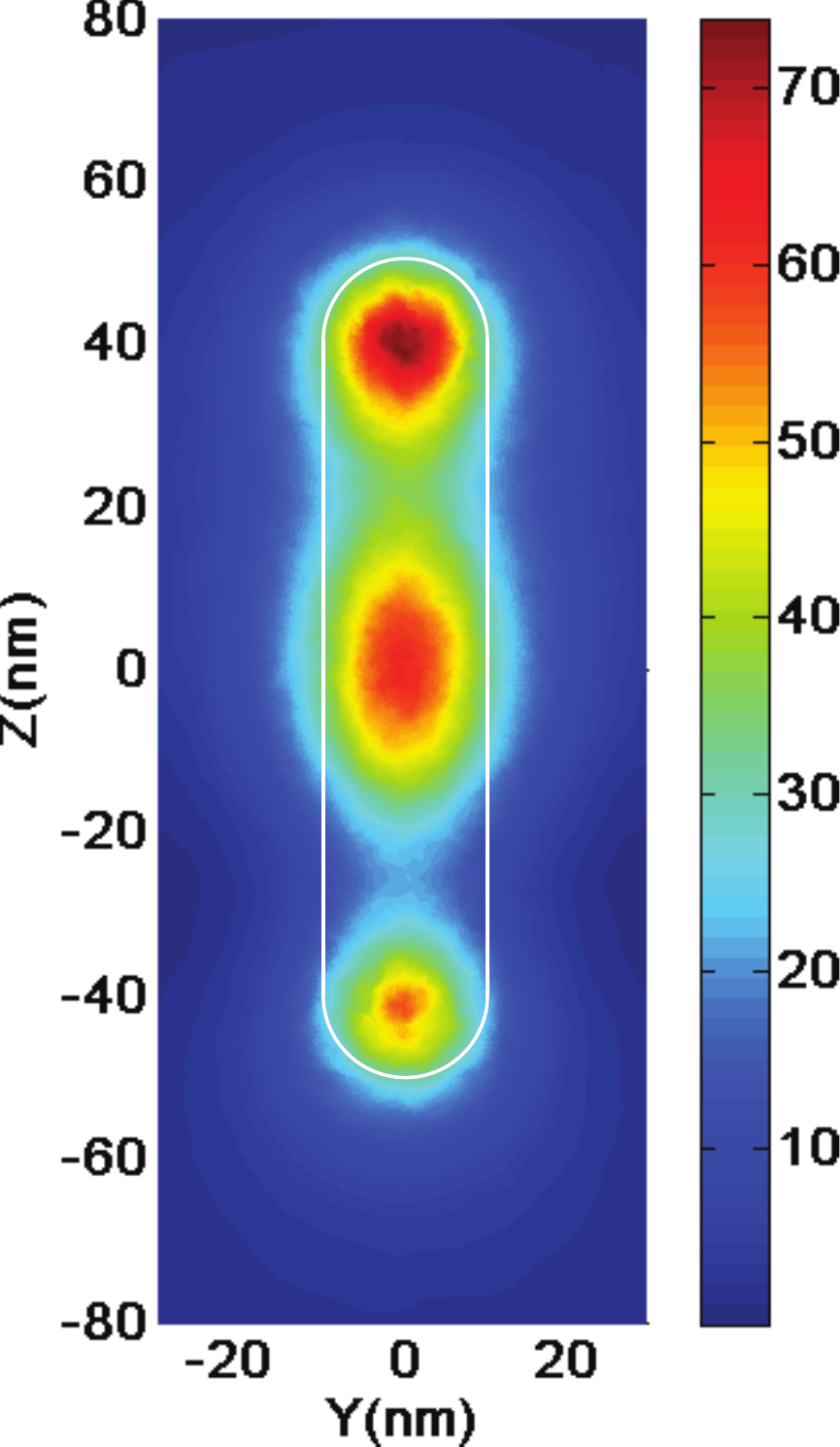,height=6cm}\label{SM_Comsol}}
\end{center}
\caption{a) Near-field intensity spectrum calculated at one extremity (point $P$ in Fig. \ref{Monomer}, located 4 nm away from the silver rod tip). 
FEM: Finite Element Method. $AR=5.33$ and $AR=1$ refer to GDT computation with cell aspect ratio $a/b=5.33$ and $a/b=1$ ($b=2.5$ nm), respectively.   
b and c) Near-field intensity distribution maps calculated 5 nm above the rod surface at the two resonance peaks. 
Calculations were performed using finite element method. The incident field is transverse magnetic (TM) 
and the incident angle is fixed at $\pi/4$. The intensity is normalized with respect to the incident intensity. The silver rod position is reported (white lines).}
\end{figure}\label{Calc_Comsol}

We first investigate the near field optical response of the silver nanorod using FEM. The PML parameters (thickness and position) were determined by benchmarking the calculation of a well-known object characterized by the exact generalized Mie theory \cite{Stout-Auger-Devilez:2008} (not shown). Considering dimer test structures, we achieve an excellent agreement with $50~\mathrm{nm}$ 
thick spherical PML located at $150~\mathrm{nm}$ from the nanostructure. We then consider the silver nanorod presented in Fig.~\ref{Monomer}. 
A simple way to determine all the supported modes of the nanostructure is to excite it with an oblique incident plane wave ${\bf E_0}$. 
We characterize  the optical near-field response by computing the electric field intensity at one extremity (point $P$ in Fig. \ref{Monomer}).

Figure \ref{Spectres} displays the near-field intensity spectra of the silver nanorod ($L=100$ nm and $D=20$ nm). 
We observe two surface plasmon polariton (SPP) resonances located around $\lambda=440~\mathrm{nm}$ and $\lambda=682~\mathrm{nm}$. 
These resonances are in agreement with a Fabry-Perot resonator description \cite{Ditlbacher2005,Novotny:2007,Cubukcu-Capasso:2009}. 
The effective indices of the SPP guided along a 20 nm silver nanowire are $\mathrm{n}_{\mathrm{eff}}=2.76$ and $\mathrm{n}_{\mathrm{eff}}=3.98$ 
at $\lambda=682~\mathrm{nm}$ and $\lambda=440~\mathrm{nm}$, respectively. The cavity modes correspond to resonator lengths 
$L+2\delta=m\lambda/(2\mathrm{n}_{\mathrm{eff}})$ with $m=1,2,...\,$ $m$ is the mode order and $\delta$ refers to the field penetration depth into air. 
We obtain $[m=1,\delta=12~\mathrm{nm}$] and $[m=2,\delta=5~\mathrm{nm}]$ for the two first modes. Note that these values for $\delta$ reveal that high order modes are strongly confined \cite{Derom-Vincent-Bouhelier-GCF:2012}. 
Figures \ref{FM_Comsol} and \ref{SM_Comsol} represent near-field intensity maps computed at these resonances located at $\lambda=682~\mathrm{nm}$ and $\lambda=440~\mathrm{nm}$, respectively. The intensity map at $\lambda=440$ nm is not symmetric due to an interference with the incident excitation field (Fig. \ref{SM_Comsol}). The apparent symmetry observed at   
$\lambda=682$ nm is simply due to the fact that the amplitude of the excited mode is stronger than the incident field (Fig. \ref{FM_Comsol}). 

We now apply the formalism of Green's method presented in the previous section. The silver nanorod is discretized with 
rectangular parallelepipeds of dimensions $a \times b \times c$. We set $b$ and $c$ equal to $2.5~\mathrm{nm}$ in the following, while varying the parameter 
$a$ from $a=2.5~\mathrm{nm}$ to $a=20~\mathrm{nm}$ as shown in Fig. \ref{Calc_Green}. The two hemispherical caps are discretized with $a=b=c=2.5~\mathrm{nm}$ 
for each case. Only the central cylindrical part of the nanorod is meshed with different $a$ values. 
We use a transitory meshing area between the cap and the central part. 
The background is not discretized contrary to FEM calculations. 

\begin{figure}[ht]
\begin{center}
  \subfigure[$\mathrm{AR=1}$]{\epsfig{figure=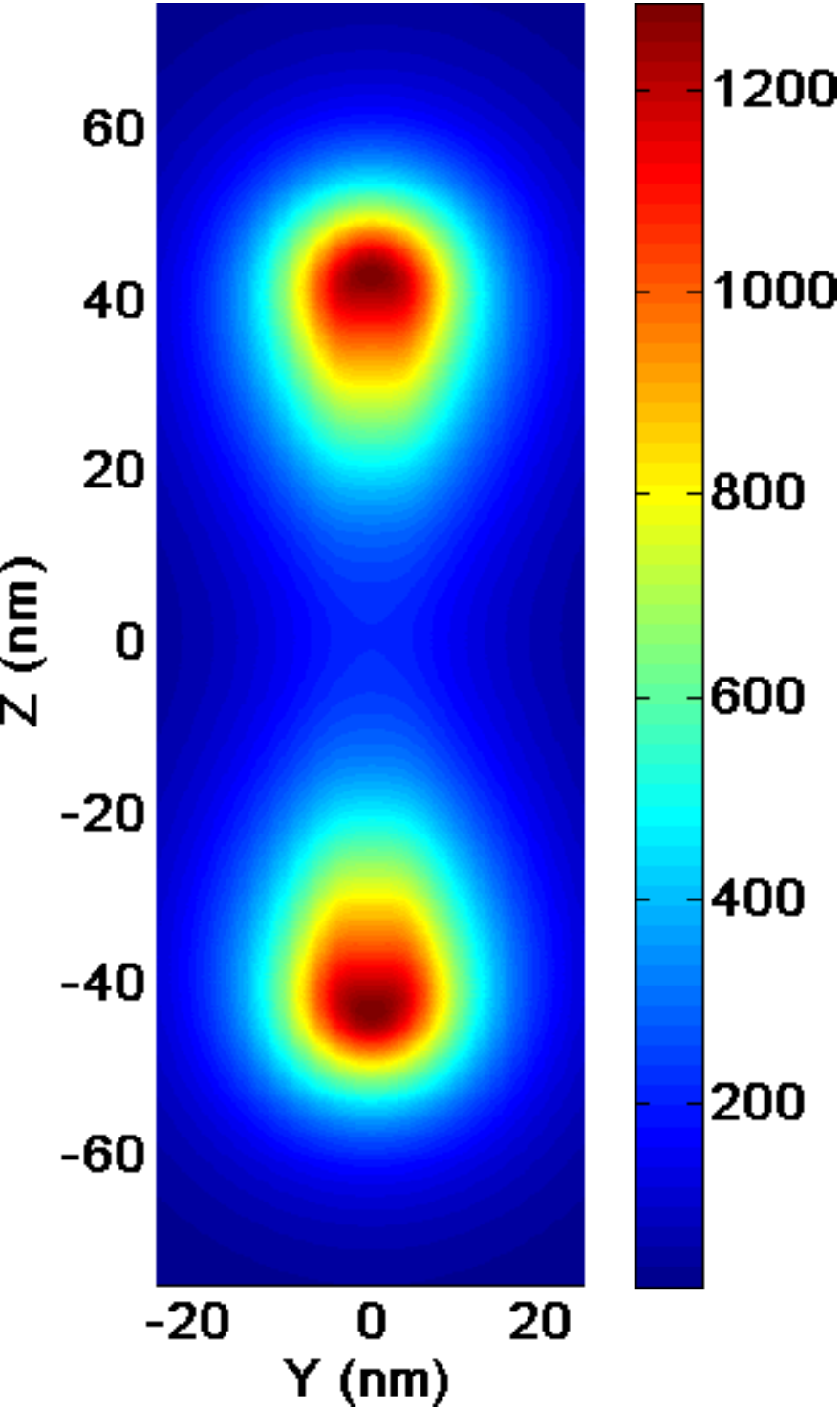,height=6cm}\label{FM_Green1}}\quad
  \subfigure[$\mathrm{AR=2}$]{~~~~\epsfig{figure=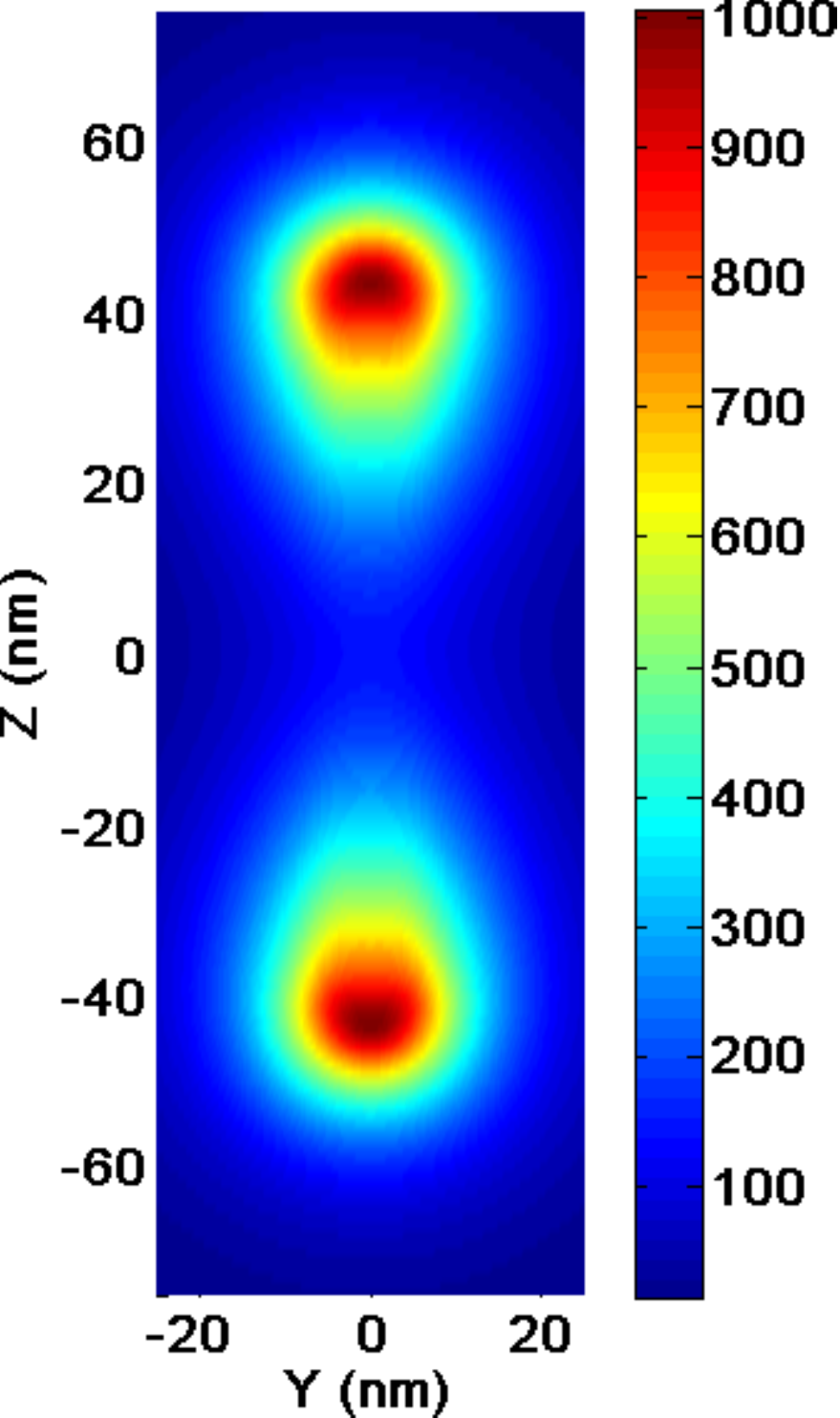,height=6cm}\label{FM_Green2}}\quad
  \subfigure[$\mathrm{AR=5.33}$]{~~~~
  \epsfig{figure=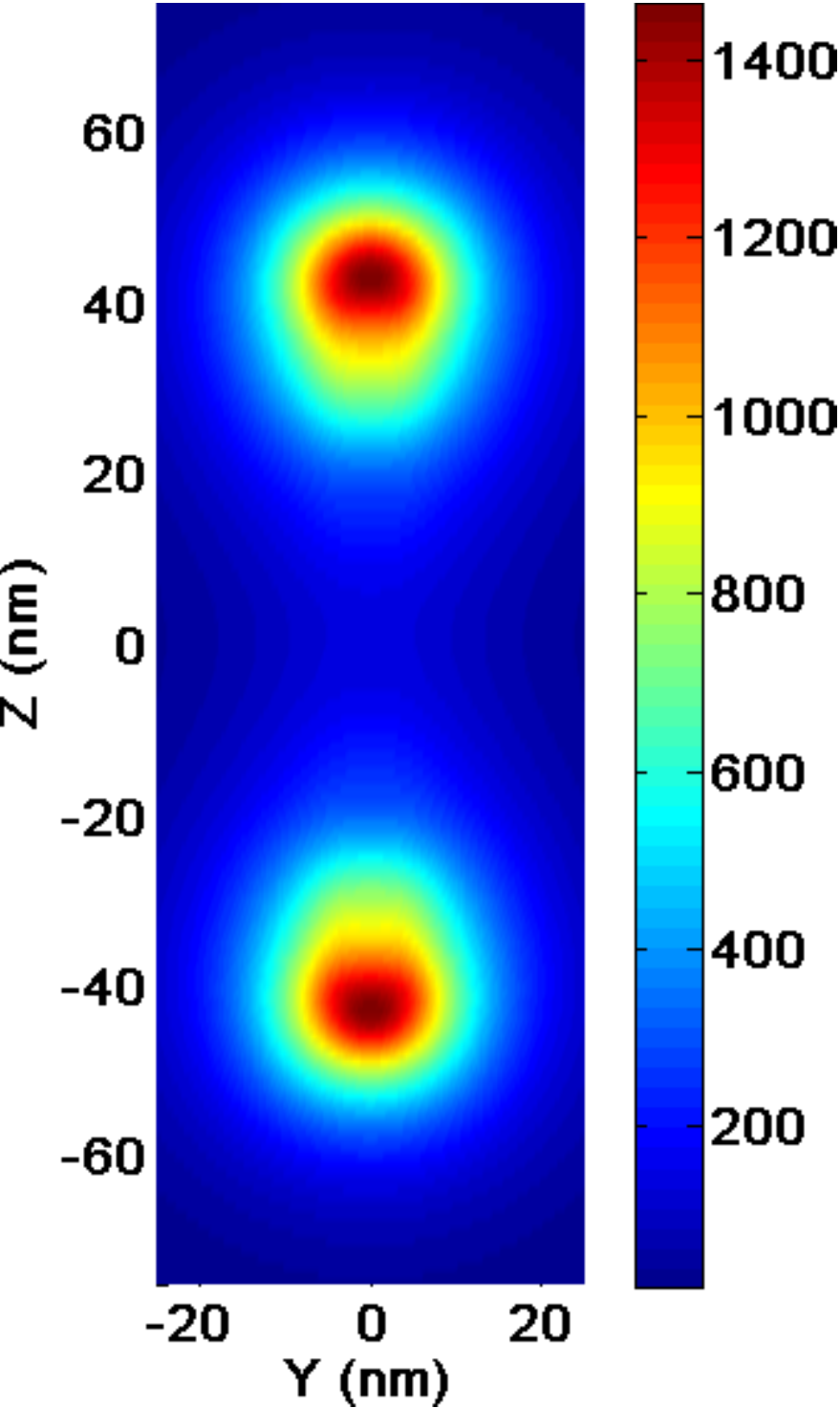,height=6cm}\label{FM_Green6}}\quad
  \subfigure[$\mathrm{AR=8}$]{\epsfig{figure=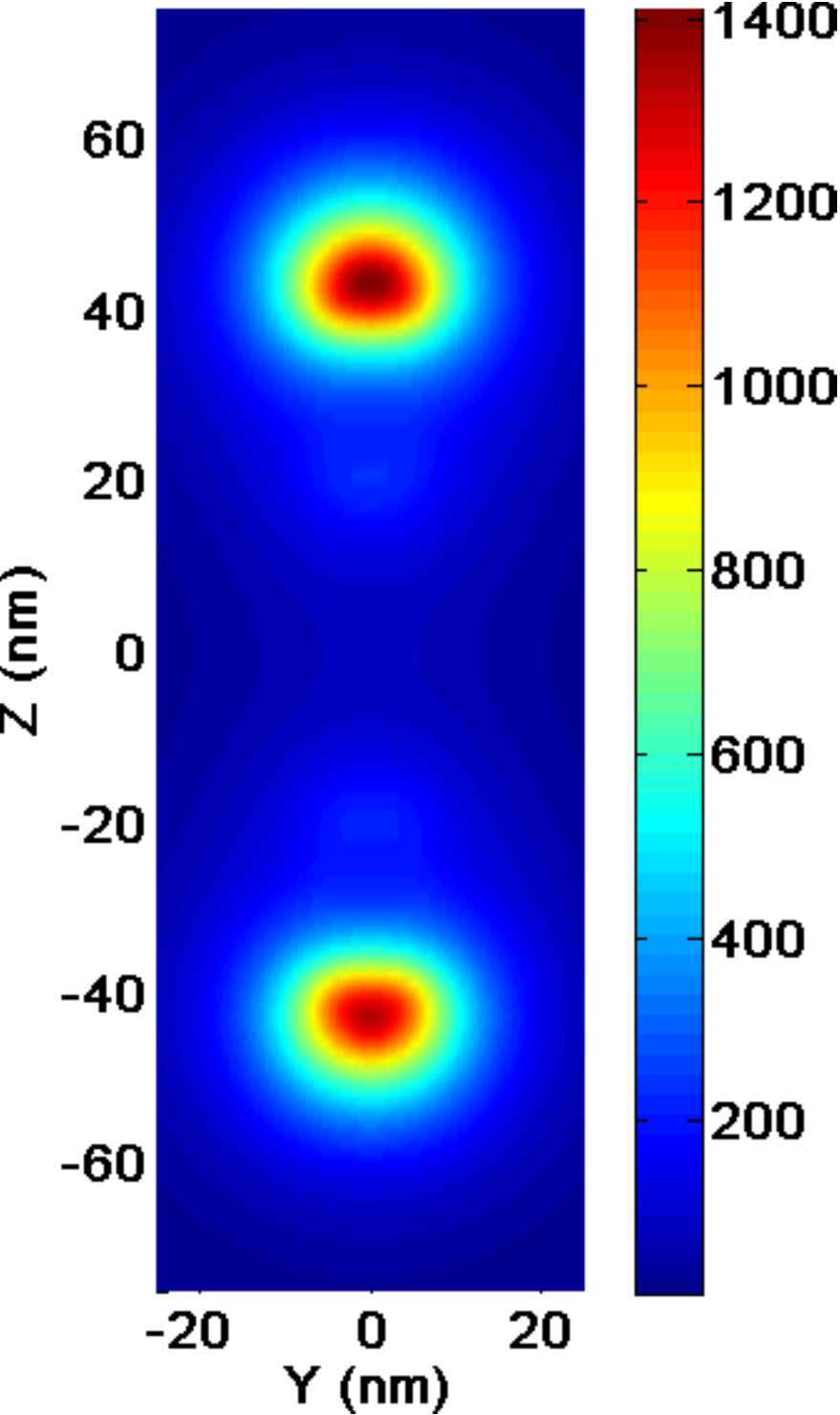,height=6cm}\label{FM_Green4}}\quad
\end{center}
\caption{Near-field intensity distribution calculated 5 nm above the rod surface at the wavelength $\lambda=682~\mathrm{nm}$. 
The hemispherical caps are discretized with cubic cells of length 2.5 nm whereas the central part is discretized with rectangular 
parallelepipeds of dimensions $a \times b \times c$ with $b=c=2.5~\mathrm{nm}$ and different aspect ratio $AR=a/b$. $a$ varies from 
$a=2.5~\mathrm{nm}$ (AR=1) to $a=20~\mathrm{nm}$ (AR=8). Calculations performed using GDT.}
\label{Calc_Green}
\end{figure}

The near-field spectrum is calculated in Fig. \ref{Spectres} for two mesh aspect ratios, namely $AR=1$ ($a=2.5~\mathrm{nm}$) and $AR=5.33$ ($a=13.3~\mathrm{nm}$).  
We observe an excellent agreement with the FEM calculated spectrum for the finer discretization, except for a peak near $\lambda=730 ~\mathrm{nm}$. 
This peak disappears when changing the meshing size so that it is easily detected as a numerical artefact. In case of meshing aspect ratio $AR=5.33$, 
the dipolar resonance ($m=1$) is in agreement with the FEM computation. The $m=2$ resonance is however slightly blue shifted.  
We represent the calculated intensity at $\lambda=682~\mathrm{nm}$ in Fig. \ref{Calc_Green} for mesh aspect ratio ranging from 1 to 8.  
When the aspect ratio of elementary cells is close to the one of the full object 
(Fig. \ref{FM_Green6}, $AR\approx 5$) we can reach a very good agreement with the result obtained by the finite element method (Fig.\ref{FM_Comsol}). 

\begin{table}[ht]
\centering
\begin{tabular}{|c|c|c|c|c|}
  \hline
  Aspect ratio & 1 & 2 & 5.33 & 8\\
  \hline
  Number of cells & 1944 & 1112 & 696 & 488 \\
   \hline
  Computational time & 3h13min & 1h43min & 1h04min & 48min \\
   \hline
\end{tabular}
\caption{Summary of computational time and the number of cells discretization according to the aspect ratio of the elementary cell. {Computations performed on a  Intel X5650 (2.66 GHz) processor.}}
\label{Table1}
\end{table}

Table \ref{Table1} indicates the number of meshing elements and the computing time for the different aspect ratios considered here. The computing time linearly increases with the cell number $N$ since the limiting operation is the numerical integration of the free-space Green's tensor over the cuboidal cell (instead of an increasing time as $N^3$ when the limiting factor is the resolution of the self consistent Dyson's equation \ref{eq:DysonDisc}).

{Apart from the finest meshing, we obtain the better result for a meshing shape concomitant with the full object aspect ratio (Fig. \ref{FM_Green6}).  We attribute this to the fact that the resonance position of an elementary mesh is close to the object resonance. 
So the optical response of the whole structure is qualitatively described by the meshing and the object discretization refines the modelisation with a scaling law improvement. Figure \ref{fig:Spectre_NbCel} represents the near-field intensity when considering a single cell (AR=5.33), 27 cells (AR=5.33) or the full object (AR=5). Although the localized plasmon resonance of a silver spherical or cubic nanoparticle is around $\lambda\approx 350$ nm, the elongated cell presents a strong red shifted resonance around $\lambda=920$ nm and  the resonance quickly converges to its final value as the object shape is built up. }

\begin{figure}
\includegraphics[width=8cm]{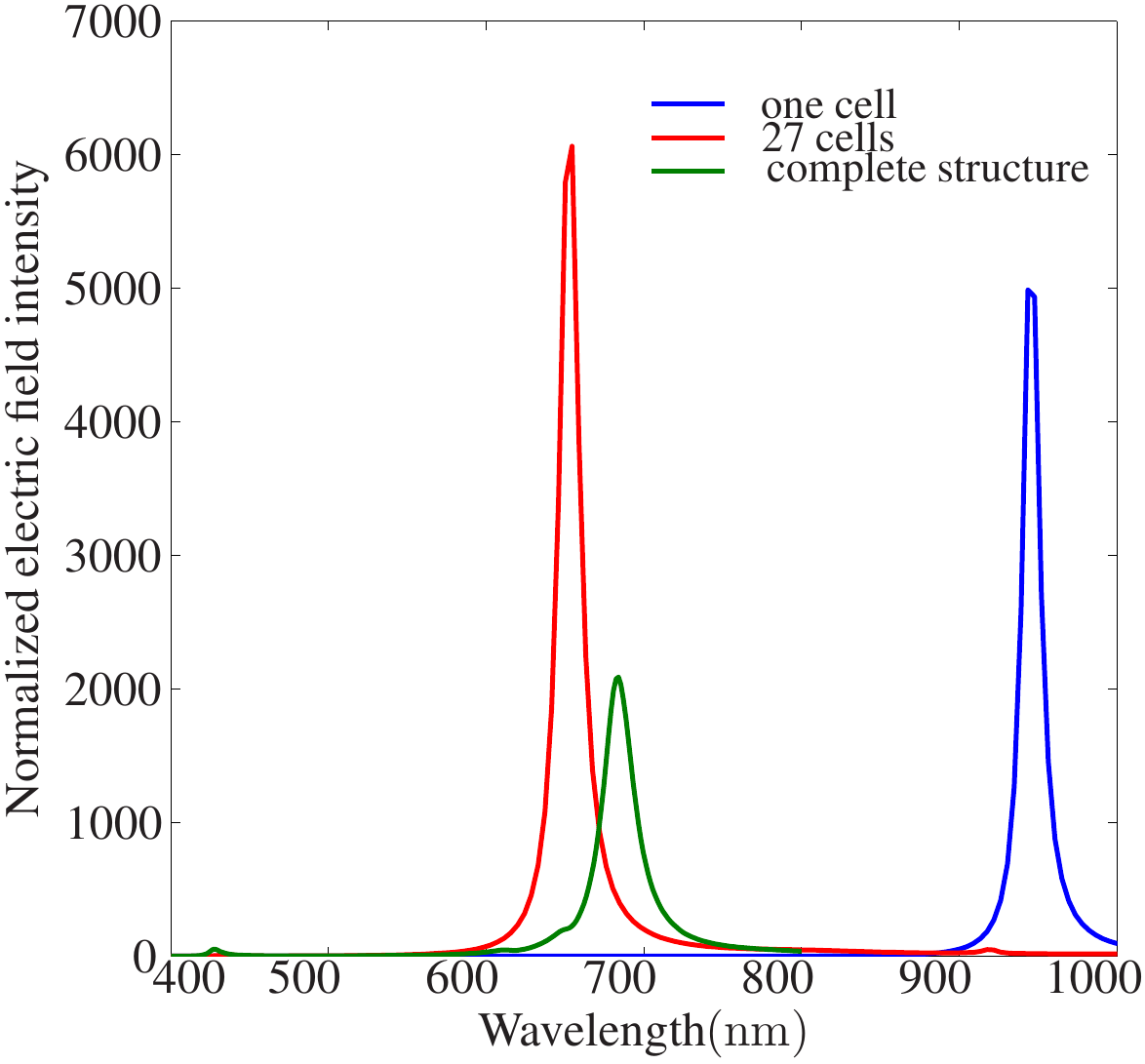}
\caption{Electric field intensity calculated 4 nm away from a single cell (blue curve, 'one cell'), a $3\times 3 \times 3$ cells object (red curve 'three cells') and the complete structure (green curve). The cell aspect ratio is AR=5.33.}
\label{fig:Spectre_NbCel}
\end{figure}

Finally, we characterize the field confinement by computing the intensity as a function of distance to the rod's extremity in figure \ref{RodMode1}. FEM and GDT with high aspect ratio meshing are in quantitative agreement. The exponential fit shows a decay distance 
$\delta=11 ~\mathrm{nm}$ in agreement with the value deduced from the Fabry-Perot resonator model ($[m=1\,, \delta=12$ nm], see above). Note that the exponential decay is a simple way to characterize the mode confinement but does not refer to the dipolar field decay \cite{Lal-Halas:2006,Deeb-Bachelot-Plain-Soppera:2010}.

\begin{figure}[ht]
\centering
    \includegraphics[width=6cm]{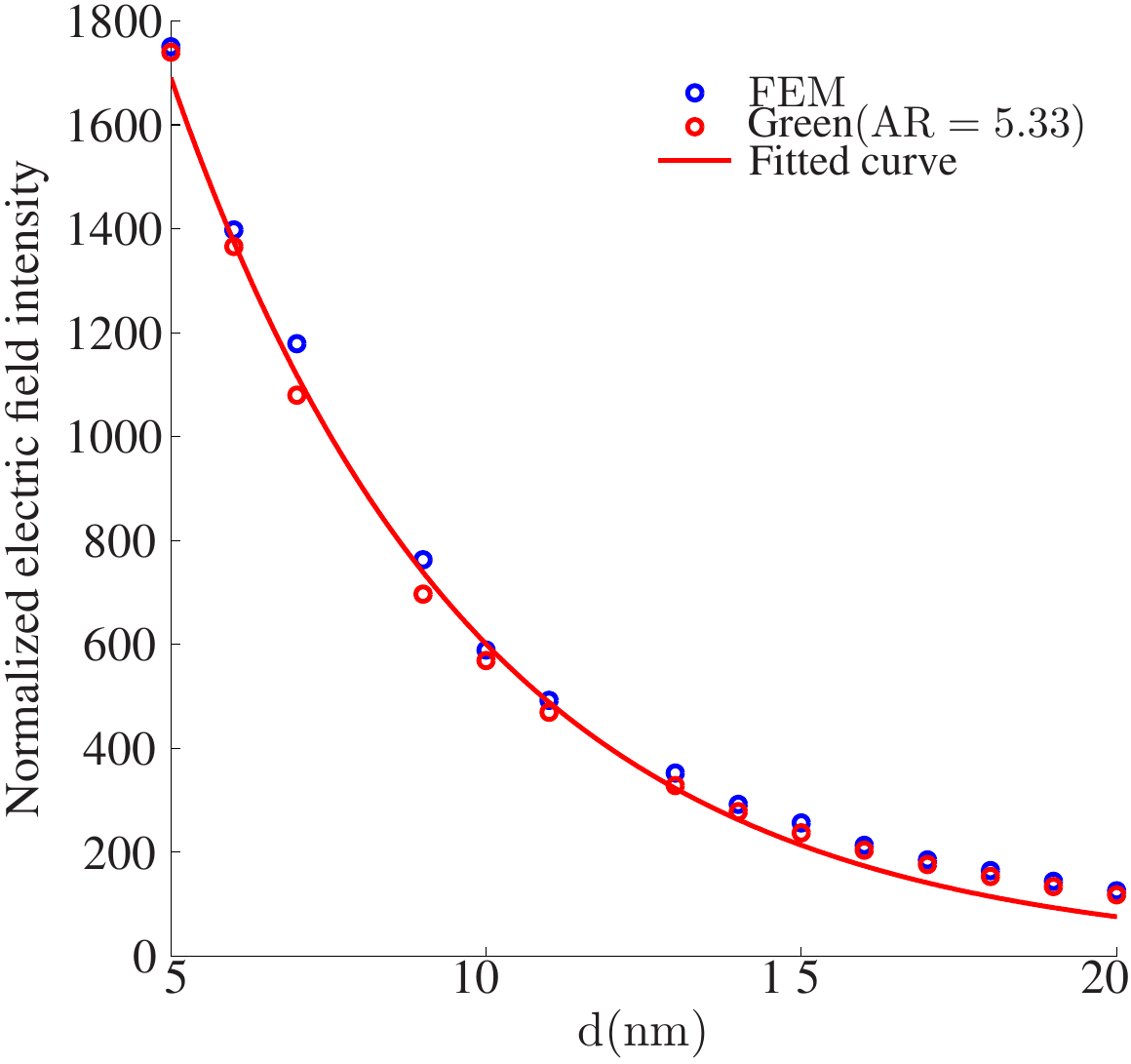}
    \caption{Normalized near-field intensity calculated as a function of the distance to the rod extremity at $\lambda=682$ nm using FEM or GDT.  
    For GDT, the hemispherical caps are discretized with cubic cell of length 2.5 nm whereas the central part is discretized with rectangular 
parallelepipeds of dimensions $a \times b \times c$ with $b=c=2.5~\mathrm{nm}$ and aspect ratio $AR=a/b=5.33$. The green curve is an exponential fit 
$I(d)=Ae^{-2d/\delta}$ with $A=4653$ and $\delta=11~\mathrm{nm}$.}
\label{RodMode1}
\end{figure}

Figure \ref{Calc_GreenSM} represents the near-field intensity calculated at $\lambda=440~\mathrm{nm}$ for different meshing.  
We obtain a good agreement with the FEM-calculated  map for the finest meshing (AR=1 and AR=2). Larger mesh aspect ratios are too rough to grasp the fast spatial field variation over the rod length. 
The agreement between the FEM and the GDT maps is not fully quantitative, even for the finest meshing 
(compare the maximum intensity in Figs. \ref{SM_Comsol} and \ref{SM_Green1},\ref{SM_Green2}). Since the optical response is extremelly sensitive to the exact object shape, we attribute this difference to the slight difference of the  object shape due to the meshing. Indeed, GDT relies on a rectangular meshing so that we expect higher scattering on the edges.  The corresponding object roughness is however of the order of $\mathrm{rms} \approx 1~\mathrm{nm}$ so that the considered meshing is sufficient to describe object obtained within the state of the art of nanofabrication techniques. 
Considering a radius of 11 nm instead of 10 nm, we obtain a quantitative agreement between the FEM ($R=11~\mathrm{nm}$) and GDT ($R=10~\mathrm{nm}$) calculated intensities. We therefore conclude 
to the agreement between the two methods, within the nanofabrication tolerance.

\begin{figure}[ht]
\begin{center}
  \subfigure[$\mathrm{AR=1}$]{\epsfig{figure=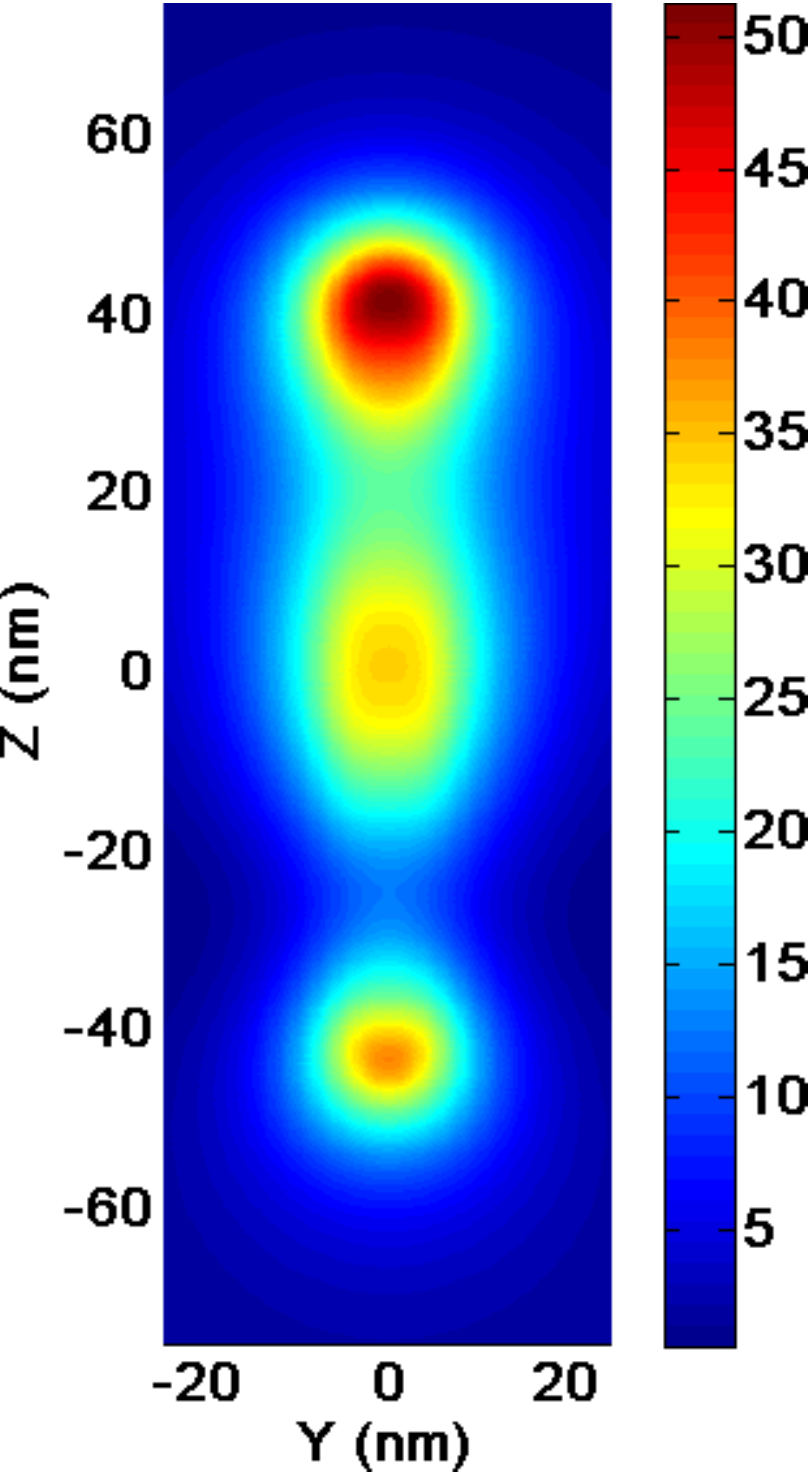,height=6cm}\label{SM_Green1}}\quad
  \subfigure[$\mathrm{AR=2}$]{~~~~\epsfig{figure=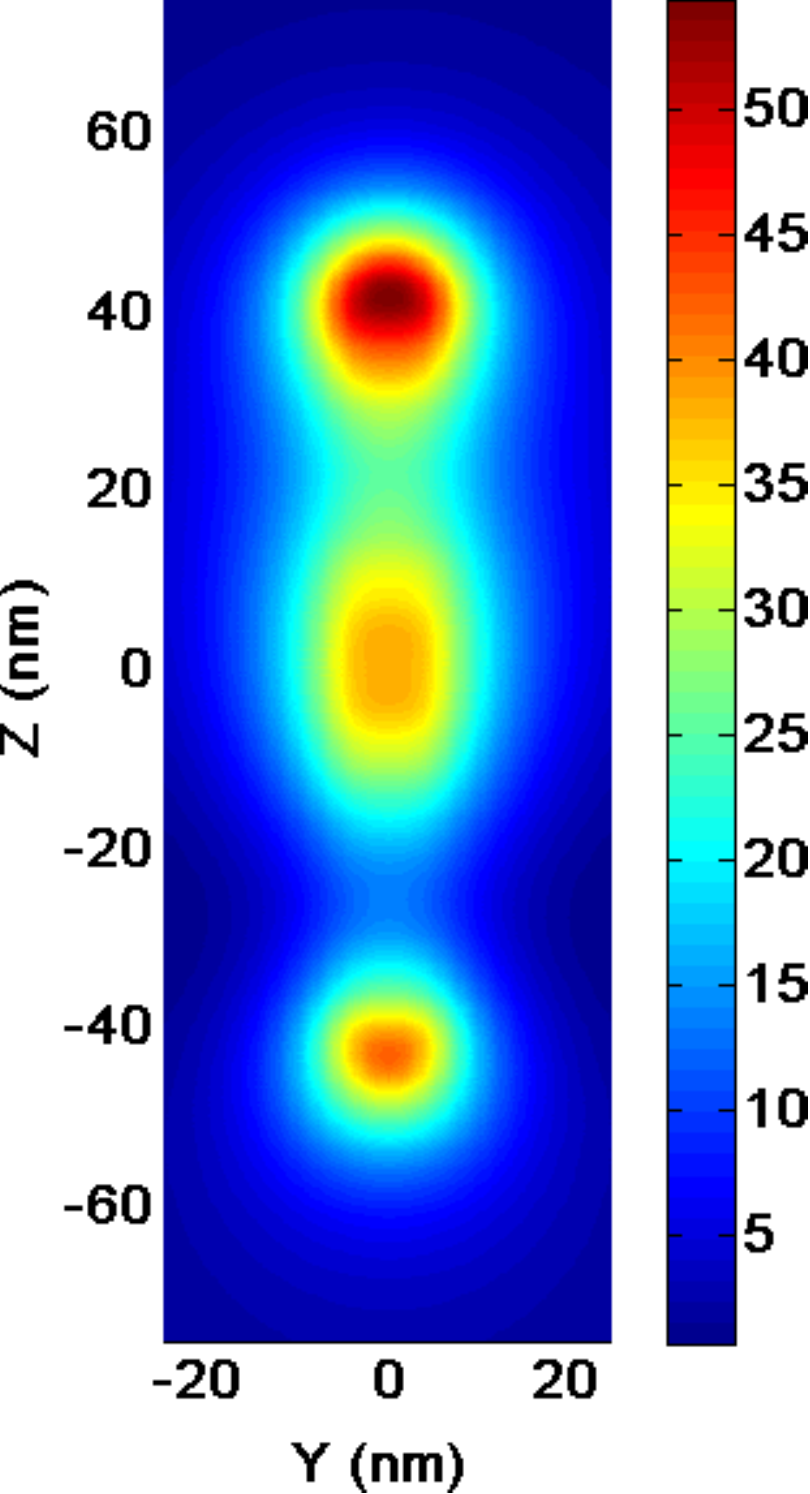,height=6cm}\label{SM_Green2}}\quad
  \subfigure[$\mathrm{AR=5.33}$]{~~~~\epsfig{figure=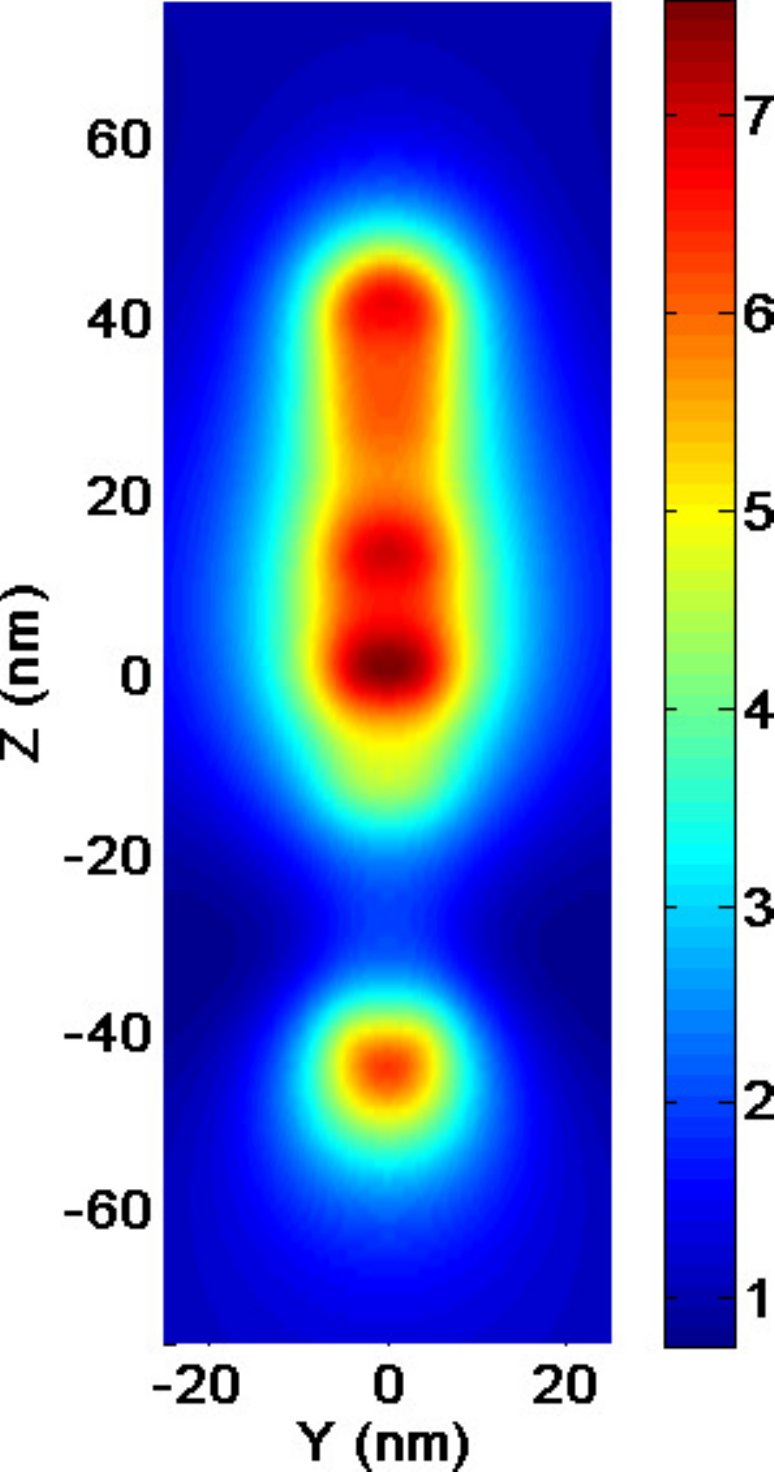,height=6cm}\label{SM_Green6}}\quad
  \subfigure[$\mathrm{AR=8}$]{\epsfig{figure=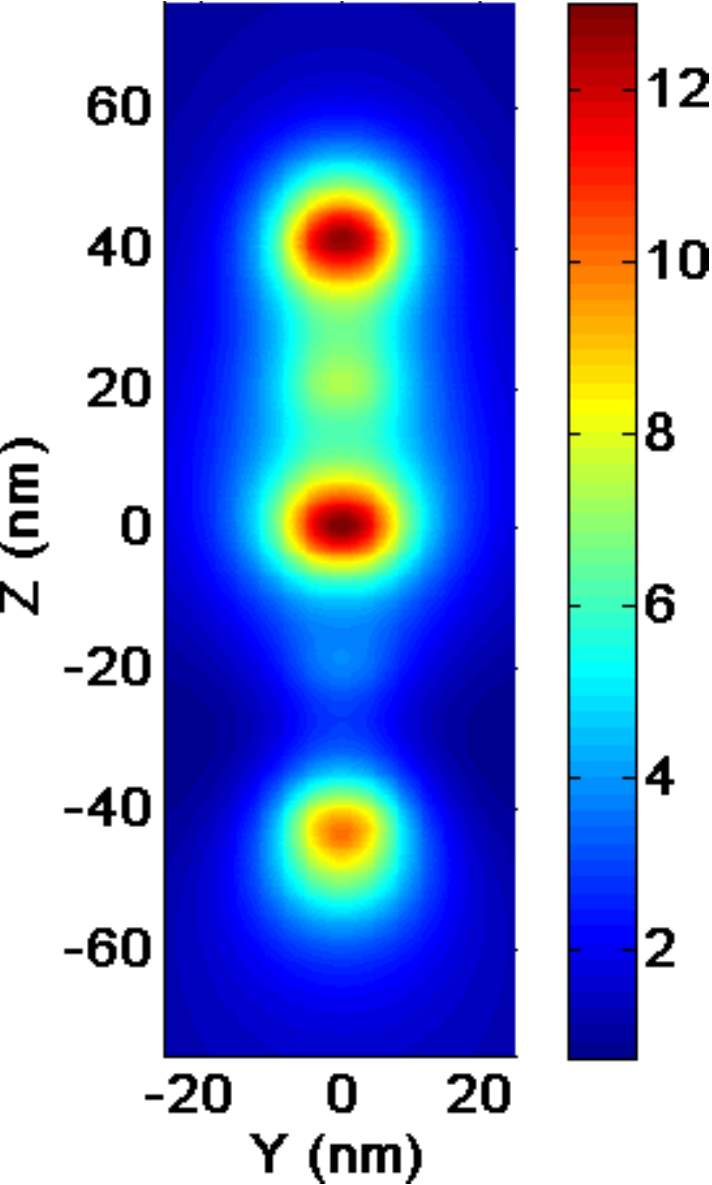,height=6cm}\label{SM_Green4}}\quad
\end{center}
\caption{Near-field intensity distribution calculated 5 nm above the rod, at the resonance of the second mode ($\lambda=440\mathrm{nm}$). Calculations performed using GDT with meshes of different aspect ratios as indicated in the figures.}
\label{Calc_GreenSM}
\end{figure}

\psection{LDOS}
\begin{figure}[ht]
\begin{center}
  \subfigure[fundamental mode]{\epsfig{figure=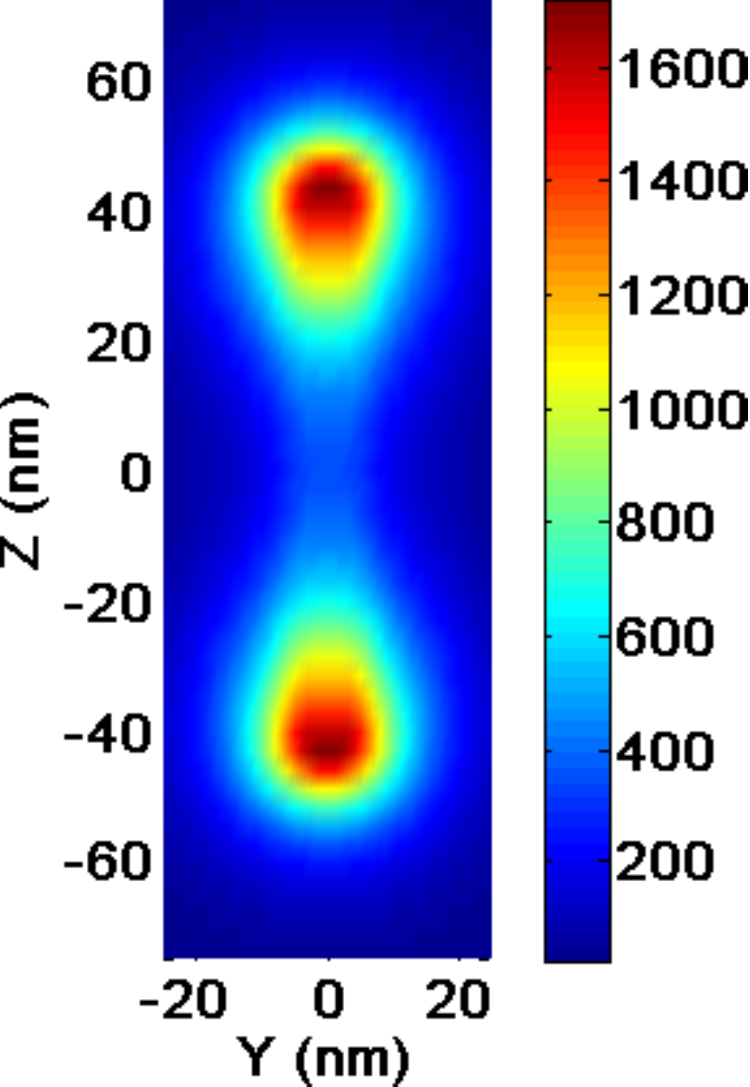,height=6cm}\label{LDOS_FM}~~~~}\quad
  \subfigure[second mode]{\epsfig{figure=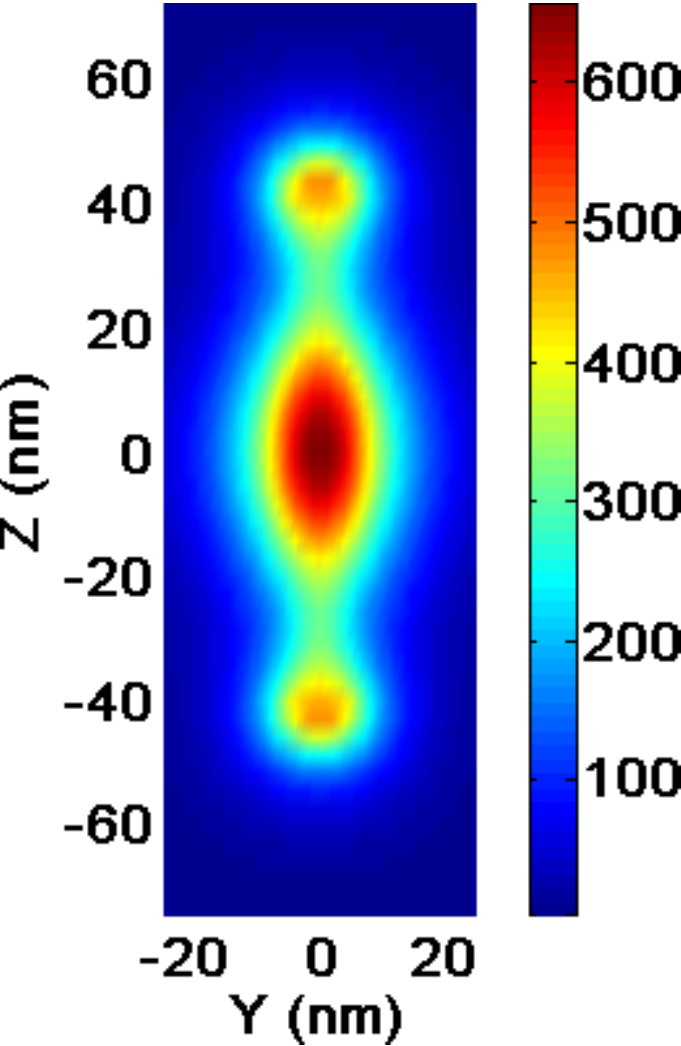,height=6cm}\label{LDOS_SM}}
\end{center}
\caption{Total LDOS map calculated 5 nm above the rod surface at the two resonance peaks. Calculations performed using GDT. LDOS is normalized with respect to its free-space value.}
\label{Calc_LDOS}
\end{figure}

So far, we discussed the optical response of the metal nanorod to a plane wave excitation. This allows to detect the plasmon resonance and estimate the filed profile of the mode. However, LDOS map better describes the mode since it is an intrinsic properties of the system. The LDOS is also a key quantity to interprete electron energy loss spectroscopy  (EELS) \cite{GarciaAbajo-Kociak:2008a,Bouhelier-ColasdesFrancs-Grandidier:2012} and governs the decay rate of a quantum emitter in confined system \cite{Barnes:1998}. It can also be manipulated at the nanoscale, opening promising perspectives to realize original nano-optical devices \cite{Viarbitskaya-Dujardin:2013}.

Figure \ref{Calc_LDOS} represents the LDOS calculated above the nanorod using Eq. \ref{eq:LDOS} at the two resonances wavelength. As expected, the LDOS maps are symmetric and characterizes the suported modes, with  a number $m$ of node in agreement with the Fabry-Perot resonator description \cite{Imura-Okamoto:2005}.  
The normalized LDOS amplitude of the first mode (up to $\approx 1700$) is comparable to the normalized electric field intensity calculated for a plane wave excitation (Fig. \ref{FM_Green1}, up to $\approx 1400$). This again reveals the efficient excitation of this mode by a plane wave. On the contrary, the second mode is weakly excited by a plane wave, even at oblique incidence (compare the LDOS magnitude, up to $\approx 600$ in Fig. \ref{Calc_LDOS}b and the electric intensity amplitudes, about ten times less in Figs. \ref{Calc_GreenSM}a and \ref{Calc_LDOS}b).

\psection{Effect of the substrate}
Finally, we discuss the effect of the substrate on the optical properties. This is an important point since plasmonic nanostructures are generally supported on a substrate. The presence of the substrate red shifts the resonance peak \cite{Weeber-Girard-Dereux-Krenn-Goudonnet:1999} so that it has to be taken into account. However, it could lead to difficult FEM numerical implementation since the field radiatively leaks into the substrate (most refringent medium) so that it has to be sufficiently discretized and requests important memory resources. On the opposite, GDT easily takes into account the presence of the substrate without increasing the memory cost and with a minor increase of the computing time. In the present case where the Green's dyad has to be integrated over the mesh surface, it is advantageous to work within the dipole image (quasi-static) approximation that leads to an excellent agreement with the exact retarded description for nanostructures laying on a glass substrate \cite{GayBalmaz-Martin:2000}. Let us consider a dipole ${\bf p}_0=(p_x,p_y,p_z)$ at position ${\bf r}_0=(x_0,y_0,z_0)$  above a substrate of dielectric constant $\epsilon_{sub}$. The effect of the substrate on the dipole radiation is described by an image dipole ${\bf p}_{im}$ at ${\bf r}_{im}=(x_0,y_0,-z_0)$ but in 
the homogeneous background of dielectric constant $\epsilon_B$. If the observation point ${\bf r}$ is above the substrate, the image dipole to consider writes  
${\bf p}_{im}=(\epsilon_{sub}-\epsilon_B)/(\epsilon_{sub}+\epsilon_B)(-p_x,-p_y,p_z)$. If the observation point is below the substrate, the image dipole to consider writes  
${\bf p}_{im}=2\epsilon_{sub}/(\epsilon_{sub}+\epsilon_B){\bf p}_0$. Since the Green's dyad expresses the field scattered by a dipolar source, the integrated formulation is easily deduced within the image dipole approximation. For instance, 

\begin{eqnarray}
{G_{sub,xx}^{int}}({\bf r},{\bf r_0})={G_{0,xx}^{int}}({\bf r},{\bf r_0})-\frac{\epsilon_{sub}-\epsilon_B}{\epsilon_{sub}+\epsilon_B} {G_{0,xx}^{int}}({\bf r},{\bf r_{im}}) \,,
\label{eq:GintSubxx}
\end{eqnarray}
for an observation point ${\bf r}$  above the substrate.

\begin{figure}[ht]
\centering
    \includegraphics[width=8cm,angle=0]{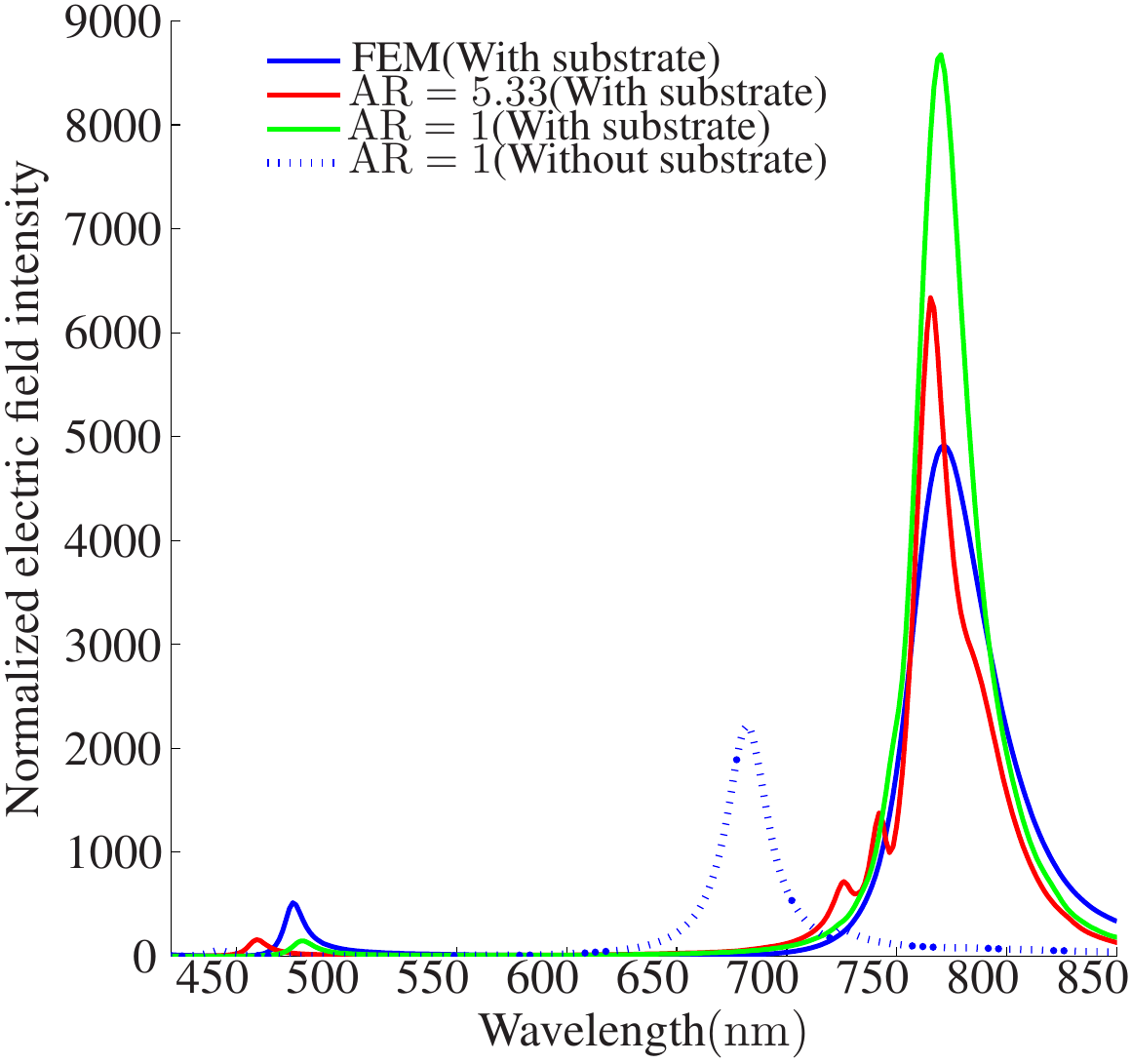}
    \caption{Near-field intensity spectrum calculated at one silver rod extremity (point $P$, located $4~\mathrm{nm}$ away from the silver rod tip). FEM: Finite Element Method. $AR=5.33$ and $AR=1$:  GDT with cell aspect ratio $a/b=5.33$ and $a/b=1$ (with $b=c=1.6$ nm), respectively. The rod is on a glass substrate of dielectric constant $\epsilon_{sub}=2.25$. The dotted curve represents the spectrum in absence of the substrate for comparison.}
\label{SpecSub}
\end{figure}

The near-field spectrum of the rod deposited on a glass substrate is calculated in figure \ref{SpecSub} using GDT.  
We observe a strong shift of the resonances to the red. 
The resonance shift from $\lambda=682~\mathrm{nm}$ to $\lambda=770~\mathrm{nm}$ for the fundamental mode and $\lambda=440~\mathrm{nm}$ to $\lambda=476~\mathrm{nm}$ for the second order mode. 
The corresponding LDOS maps are represented in figure \ref{Calc_LDOS_Sub}. We check that the mode profiles still correspond to the m=1 and m=2 SPP modes. 

We also plot  the near-field spectrum calculated using the FEM (Fig. \ref{SpecSub}). We observe good agreement with our GDT calculations except a lower intensity. We again attribute this to the 
extremelly sensitivity of the optical near-field response to the object shape. By considering a rod radius R=11 nm, FEM calculations lead to intensities comparable with GDT data. In addition, we would like to mention that the introduction of the PML layer is very critical for FEM in presence of the substrate. We observe strong variations of the calculated intensity with the PML center position. Best results are obtained when centering the PML at the rod (scatterer) center. If the PML is centered {\it e.g.} at the glass/air interface (that corresponds to a 10 nm shift only), the calculated intensity is divided by 2 (not shown). Therefore, comparison with the GDT allows to determine optimized PML parameters (size and position) for FEM before considering more complex shapes.

\begin{figure}[ht]
\begin{center}
 \subfigure[fundamental mode (m=1)]{\epsfig{figure=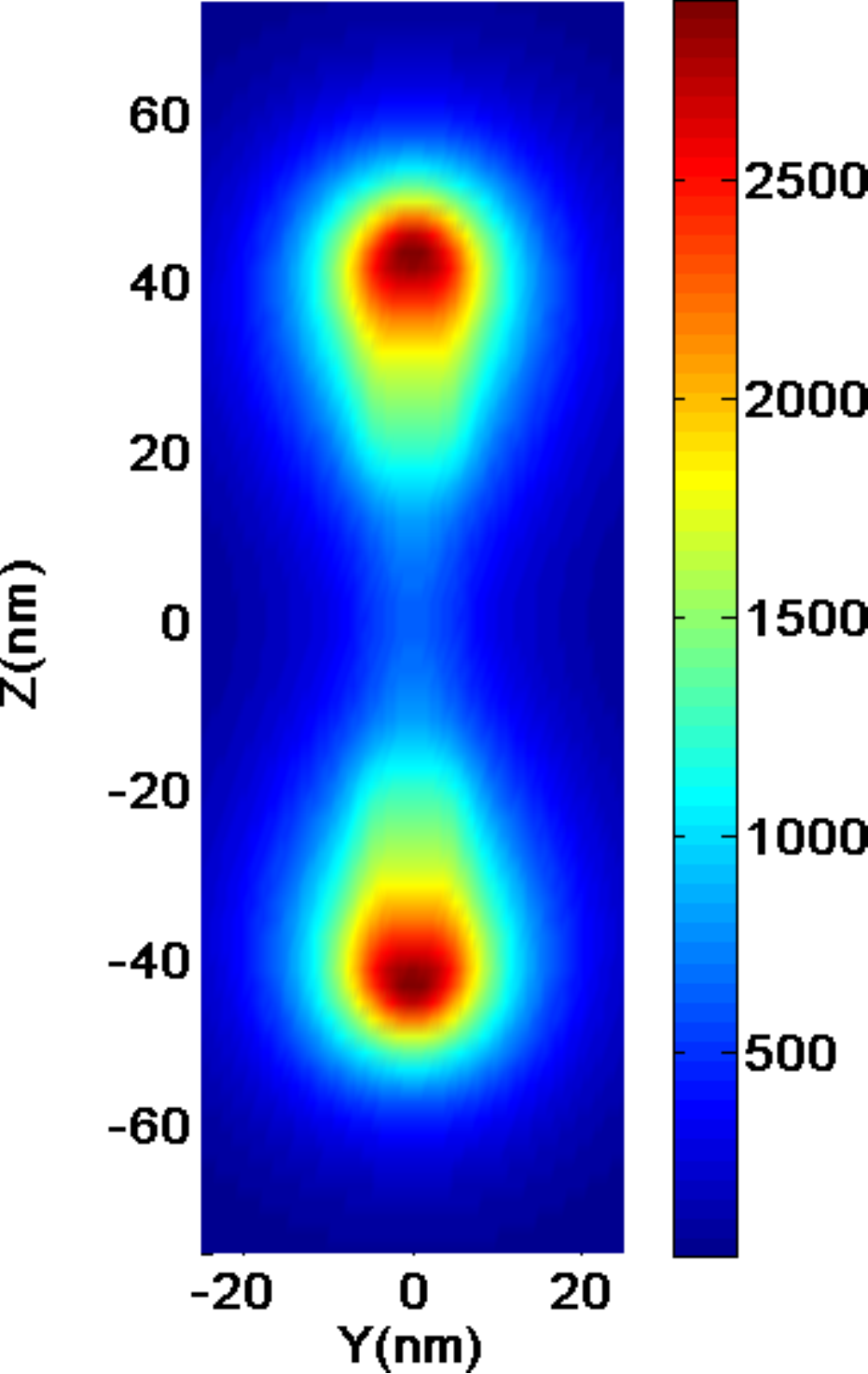,height=6cm}\label{LDOS_Sub_FM}~~~~}\quad
  \subfigure[second mode (m=2)]{\epsfig{figure=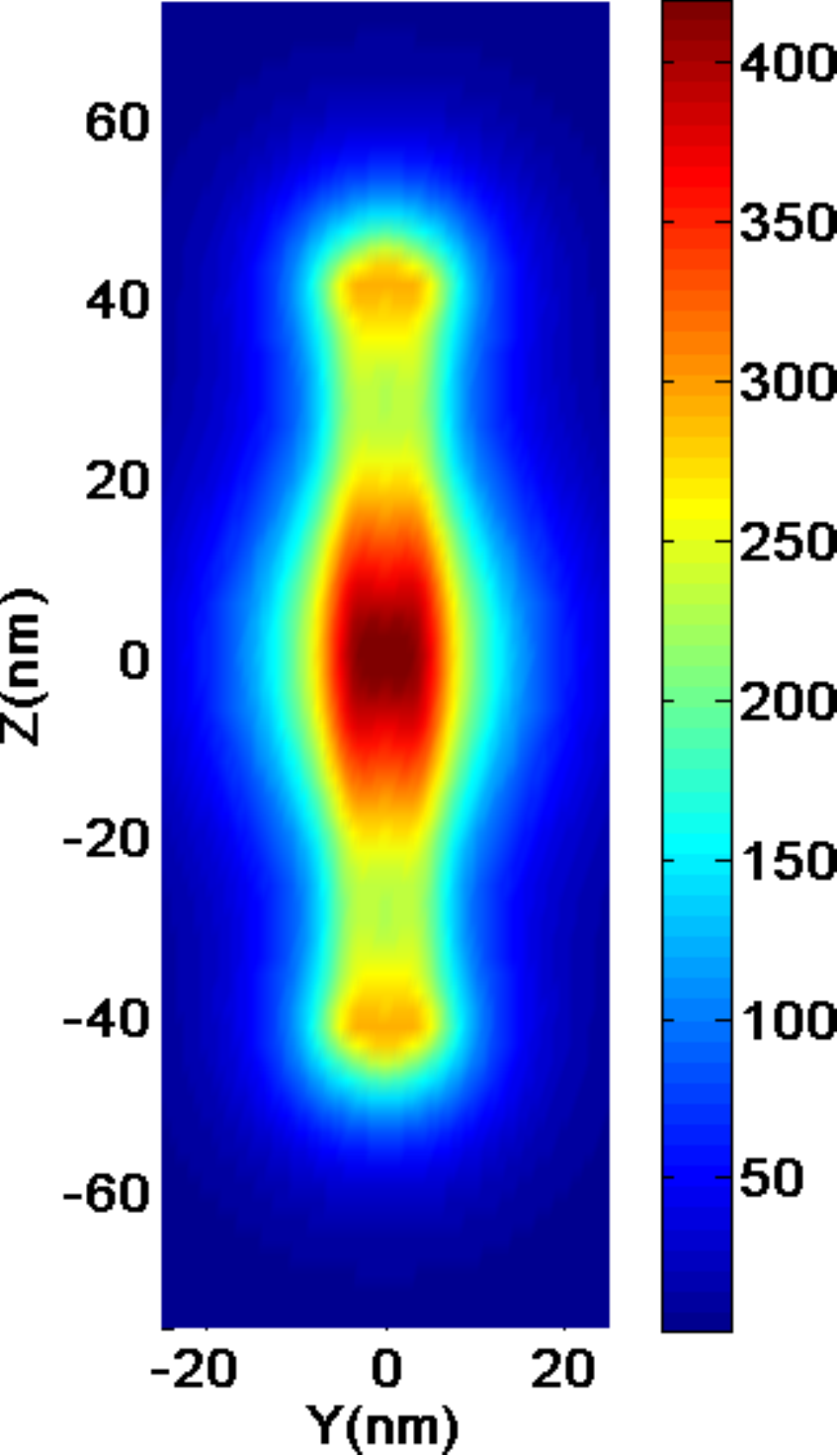,height=6cm}\label{LDOS_Sub_SM}}
\end{center}
\caption{Total LDOS map calculated 5 nm above the rod surface at the two resonance peaks. Calculations performed using GDT. 
The rod is on a glass substrate of dielectric constant $\epsilon_{sub}=2.25$.}
\label{Calc_LDOS_Sub}
\end{figure}

\psection{Conclusion}
\label{sect:CCL}
In summary we used a dyadic Green technique based on a cuboidal meshing to compute the electric near-field and the LDOS near a silver nanorod. 
The fast variations of the electric field in the metallic nanostructures are well described by a numerical integration of the Green's tensor over elongated meshes. Moreover, the computing efforts are reduced by transforming the volume integration over a mesh to a surface integral. The efficiency and accuracy of this method are verified 
by comparing our results to those obtained by using finite element method. This is a confirmation of our technique, but reciprocically helps to determine the best parameters for FEM (notably PML).  
Green's dyad technique presented in this paper is very convenient, especially in the case of elongated structure (with high aspect ratio). 
Moreover, since the meshing has to be increased at positions where the field fastly varies, this method could be advantageously combined with top-down extended meshing algorithms \cite{Alegret-Kall-Johansson:2007}.
In addition, the presence of a substrate is easily taken into account, demonstrating the versatility and efficiency of the method. Since the method relies on an numerical integration of the Green's dyad over a mesh surface, more complex 
meshing (e.g. tetrahedral) could also be considered, allowing for a better description of the object shapes.  Finally, as soon as the Green's tensor of the whole structure is calculated, 
every electromagnetic responses of the system are easily determined (electric and magnetic near and far field, LDOS, optical forces, EELS response, ...).

\psection{Ackwnoledgements}
O.D. acknowledges a stipend from the French program Investissement d'Avenir (LABEX ACTION ANR-11-LABX-01-01).  
The research leading to these results has received funding from the  Agence Nationale de la Recherche
(grants PLACORE ANR-13-BS10-0007 and HYNNA ANR-10-BLAN-1016 ) and 
the European Research Council under the European Community's Seventh Framework
Program FP7/2007-2013 (ERC SWIFT - Grant Agreement 306772).
Calculations were performed using DSI-CCUB resources (Universit\'e de Bourgogne). 

\end{paper}

\end{document}